\documentstyle[sprocl]{article}

\input{psfig}

\def\be{\begin{equation}}
\def\ee{\end{equation}}
\def\bea{\begin{eqnarray}}
\def\eea{\end{eqnarray}}

\begin{document}

\title{DARK MATTER, DARK ENERGY, AND FUNDAMENTAL PHYSICS}

\author{Michael S. TURNER}

\address{Astronomy \& Astrophysics Center, Enrico Fermi
Institute\\The University of Chicago\\5640 So. Ellis Avenue,
Chicago, IL~~60637-1433, USA}
\vskip 0.1in
\address{NASA/Fermilab Astrophysics Center, Box 500\\
Fermi National Accelerator Laboratory\\
Batavia, IL~~60510-0500, USA}
\vskip 0.05in
\address{E-mail: mturner@oddjob.uchicago.edu}


\maketitle\abstracts{More than sixty years ago Zwicky made the
case that the great clusters of galaxies are held together by the
gravitational force of unseen (dark) matter.
Today, the case is stronger and more
precise:  Dark, nonbaryonic matter accounts for $30\% \pm 7\%$
of the critical mass density, with baryons (most of which are
dark) contributing only $4.5\%\pm 0.5\%$ of the critical density.
The large-scale structure that exists in the Universe indicates
that the bulk of the nonbaryonic dark matter must be cold
(slowly moving particles).  The SuperKamiokande
detection of neutrino oscillations shows that particle
dark matter exists, crossing an important threshold.
Over the past few years a case has developed for a dark-energy problem.  
This dark component
contributes about $80\%\pm 20\%$ of the critical density
and is characterized by very negative
pressure ($p_X < -0.6\rho_X$).  Consistent with this picture of
dark energy and dark matter are measurements of CMB anisotropy
that indicate that total contribution of matter and energy
is within 10\% of the critical density.  Fundamental physics
beyond the standard model is implicated in both the dark matter
and dark energy puzzles:  new fundamental particles (e.g., axion
or neutralino) and new forms of relativistic energy (e.g., vacuum energy
or a light scalar field).  A flood of observations will shed
light on the dark side of the Universe over the next two decades;
as it does it will advance our understanding of the Universe
and the laws of physics that govern it.
}

\section{In the Beginning ...}

The simplest universe would contain just matter.  Then, according to
Einstein, its geometry and destiny would be linked:  a high-density
universe ($\Omega_0 > 1$) is positively curved and eventually
recollapses; a low-density universe is negatively curved and expands
forever; and the critical universe ($\Omega_0 =1$) is spatially
flat and expands forever, albeit at an ever decreasing rate.

As described by Sandage, such a universe is today characterized by
two numbers:  the expansion rate $H_0 \equiv
\dot R(t_0) / R(t_0)$ and the deceleration parameter
$q_0 \equiv -\ddot R(t_0)/H_0^2R(t_0)$ where $R(t)$ is the
cosmic scale factor and $t_0$ denotes the age of the Universe
at the present epoch.  Through Einstein's equations the deceleration
parameter and density parameter are related:  $q_0 = \Omega_0 /2$.
There is a consensus that we are finally closing in on the expansion rate:
$H_0 = 65\pm 5\,{\rm km\,sec^{-1}\,Mpc^{-1}}$ (or $h=0.65\pm 0.05$).\cite{H0}
Type Ia supernovae seem to have provided the first
reliable measurement of the deceleration parameter\cite{SNe}  -- and
a surprise:  the Universe is accelerating not decelerating.
So much for a simple Universe.

We have known for thirty years our Universe is not as simple
as two numbers; it is much more interesting!  In 1964 Penzias and Wilson
discovered the cosmic microwave background radiation (CMB).  Today
the CMB is a minor component, $\Omega_{\rm CMB} = 2.48h^{-2}\times
10^{-5}$ which modifies the relationship between the density parameter
and deceleration parameter only slightly.  However, the CMB changes
the early history of the Universe in a profound way:  Earlier than about
$40,000\,$yrs the dynamics of the Universe are controlled by the energy
density of the CMB (and a thermal bath of other relativistic particles)
and not matter, with the temperature being the
most important parameter for describing the events taking place.

Not only do we live in a very interesting Universe, but also
fundamental physics is crucial to understanding its past,
present and future.  Figure 1, which summarizes the present make up of
the Universe, makes the point well:\cite{DM_SUM}  in units of the critical
density CMB photons and relic relativistic
neutrinos contribute about 0.01\%; bright stars contribute about
0.5\%; massive neutrinos
contribute more than $0.3\%$ (SuperK), but less than about 15\% 
(structure formation); baryons (total) contribute $4.5\pm 0.5\%$;
matter of all forms contributes $35\pm 7\%$;
and dark energy contributes $80\pm 20\%$.  By matter I mean particles with
negligible pressure (i.e., nonrelativistic, or in terms of a temperature,
$T\ll mc^2$); by dark energy I mean stuff with pressure whose
magnitude is comparable to its energy density but negative.

\begin{figure}
\centerline{\psfig{figure=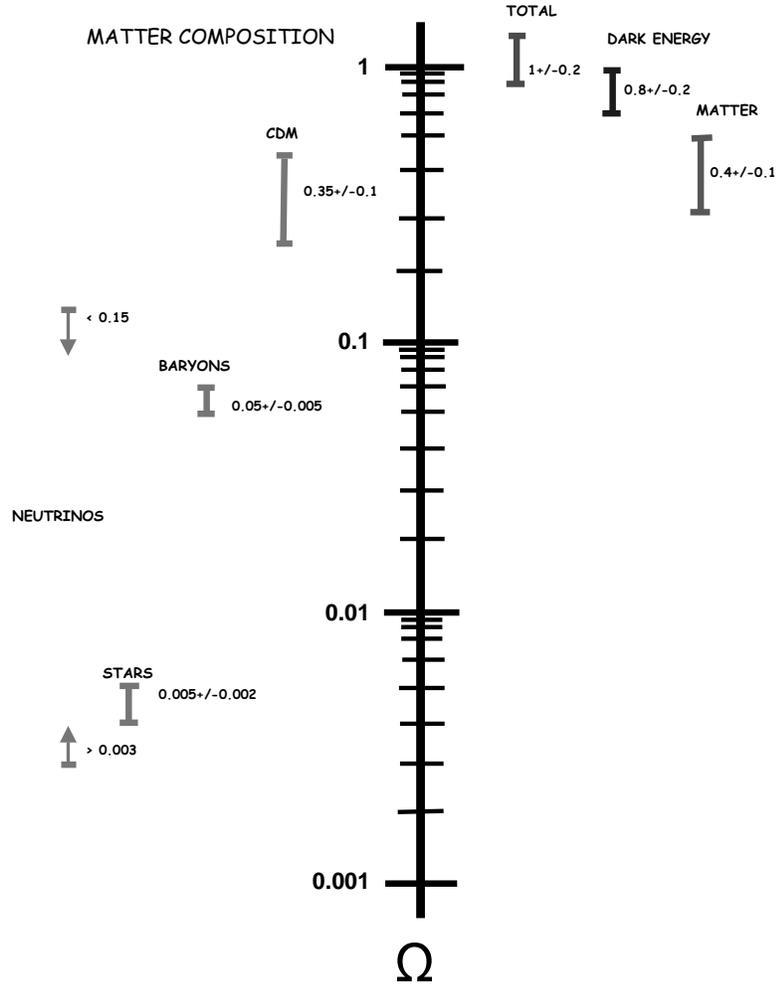,width=4.0in}}
\caption{Summary of matter/energy in the Universe.
The right side refers to an overall accounting of matter
and energy; the left refers to the composition of the matter
component.  The contribution of relativistic particles,
CBR photons and neutrinos, $\Omega_{\rm rel}h^2 = 4.170\times
10^{-5}$, is not shown.  The upper limit to mass density contributed
by neutrinos is based upon the failure of the hot dark
matter model of structure formation, and the lower limit follows from the
evidence for neutrino oscillations.
$H_0$ is taken to be $65\,{\rm km\,s^{-1}\,Mpc^{-1}}$.
}
\label{fig:omega}
\end{figure}

While cosmology is much more than two numbers, the second of Sandage's
two numbers
is still very interesting and at the heart of much of what is
most exciting today.  Allowing for a Universe with more than just
matter in it, the deceleration parameter becomes:
\begin{equation}
q_0 = {\Omega_0 \over 2} + {3\over 2}\sum_i \Omega_i w_i
\label{eq:q0}
\end{equation}
where $\Omega_0 \equiv \sum_i \rho_i /\rho_{\rm CRIT}$, $\Omega_i$ is
the fraction of critical density contributed by component $i$
and $p_i\equiv w_i \rho_i$ characterizes the pressure of component $i$
(e.g., matter, $w_i =0$, radiation, $w_i={1\over 3}$ and vacuum
energy, $w_i=-1$), and $\rho_{\rm CRIT} = 3H_0^2/8\pi G = 
1.88h^2\times 10^{-29} \,{\rm g\,cm^{-3}}$.  Note, the
energy density in component $i$ evolves as $R^{-3(1+w)}$:
$R^{-3}$ for matter, $R^{-4}$ for radiation,
and constant for vacuum energy.

The density parameter $\Omega_0$ determines the geometry of the Universe:
\begin{equation}
R_{\rm CURV} = {H_0^{-1}\over |\Omega_0 - 1|},
\end{equation}
but not necessarily its destiny.  In particular, the simple connection
between geometry and destiny mentioned earlier does not hold if
there is a component to the energy density with $w_i<-{1\over 3}$.\cite{krauss}

\subsection{Dark matter past}

The dark matter story begins with Zwicky in 1935.  He observed that
the velocities of galaxies within the great clusters of galaxies
(e.g., Coma and Virgo) are too large for the
gravity of the stars within the galaxies to hold the clusters together.
In the 1970s Vera Rubin and others\cite{IAU} measured galactic
rotation curves
(circular orbital velocity vs. radial distance from the galactic center)
using stars and clouds of neutral
hydrogen gas as test particles.
The most conspicuous feature of these rotation curves
is their flatness.  According to Newtonian mechanics this implies
an enclosed mass that rises linearly with galactocentric distance.
However, the light falls off rapidly.  Hence, the matter
that holds ordinary spiral galaxies together must be ``dark.''

In the early 1980s, a confluence of events spurred interest in
the possibility that the dark matter is exotic (nonbaryonic).
Those events included:  the growing appreciation of the deep connections
between particle physics and cosmology, a Russian experiment that
indicated the electron neutrino had a mass of around 30\,eV
(the mass needed to close the Universe for $h\sim 0.6$), and the growing
case for gap between the dark matter density needed to hold the
Universe together and what baryons can account for.
While the Russian experiment proved to be wrong, the case for nonbaryonic
dark matter grew and the inner space/outer space connection flourished.

\subsection{Dark matter present}

The case for nonbaryonic dark is now very solid and follows from
the inequality, $\Omega_M =0.35\pm 0.07 \gg \Omega_B = 0.0045\pm 0.005$.
Briefly, here is where we stand.  Big-bang nucleosynthesis provides
the best accounting of the baryons.  A precise
determination of the primeval abundance of deuterium has allowed the baryon
density to be very accurately pegged:  $\Omega_B = (0.019\pm 0.001)h^{-2}
\simeq 0.045\pm 0.005$.\cite{BBN}  From this follows the best
determination of the total matter density.

The ratio of baryons to total mass in clusters
has been determined from a sample of more than 40 clusters using x-ray
and Sunyaev-Zel'dovich measurements:  $f=(0.075\pm 0.002)
h^{-3/2}$.\cite{CBF}  (The fact that only about 15\% of the matter known
to be in clusters can be accounted for as baryons is already strong evidence
for nonbaryonic dark matter.)  Making the assumption that clusters provide
a fair sample of matter, a very reasonable assumption given their large
size, one can equate $f$ to $\Omega_B/\Omega_M$ and use the BBN value
for $\Omega_B$ to infer:  $\Omega_M = 0.35\pm 0.07$.

There is plenty of supporting evidence for this value of the mean
matter density.\cite{DM_SUM}  It comes from
studying the evolution of the abundance of clusters (with redshift),
measurements of the power spectrum of large-scale
structure, relating measured peculiar velocities to the observed
distribution of matter, and observations of the outflow of material from
voids.  Further, every viable model
for explaining the evolution of the observed structure in the Universe from
density inhomogeneities of the size detected by COBE and other CMB
anisotropy experiments requires nonbaryonic dark matter.

We have a very strong case that the bulk of the
nonbaryonic dark matter is cold dark matter (slowly moving particles).
This is based upon the many successes of the cold dark matter scenario
for the formation of structure in the Universe, as well as
the many failures of the hot dark matter scenario.  We also
have two very compelling -- and highly testable -- particle candidates:  the
axion and the neutralino.\cite{candidates}  A very light axion (mass
$\sim 10^{-6}\,{\rm eV} - 10^{-4}\,$eV) is motivated by the use of
Peccei-Quinn symmetry to solve the strong CP problem.  A neutralino of
mass 50\,GeV to 500\,GeV is motivated by low-energy supersymmetry.

On the experimental side, we now have the first evidence
for the existence of particle dark matter.  The SuperKamiokande
Collaboration has presented a very strong case for neutrino oscillations
based upon the direction dependent deficit of atmospheric muon neutrinos,
which implies at least one of the neutrinos has a mass greater
than about 0.1\,eV.\cite{SuperK}  This translates into a neutrino
contribution to the
critical density of greater than about 0.3\% (about what stars contribute).
{\em The issue is no longer the existence of particle dark matter, but
the quantity of particle dark matter.  An important threshold has been
crossed.}

There are now experiments operating with sufficient sensitivity to directly
detect particle dark matter in the halo of our own galaxy for
the two most promising CDM candidates:  axions and neutralinos.\cite{candidates}
The axion dark matter experiment
at Livermore National Laboratory is slowly scanning the favored
mass range; the DAMA experiment in Gran Sasso
and the CDMS experiment in the Stanford Underground Facility
(soon to be relocated in the Soudan Mine in Northern Minnesota)
are now probing  a part of neutralino
parameter space that is favored by theory.

\subsection{Baryonic dark matter and MACHOs}

There are actually two dark-matter problems; the second being
the discrepancy between the mass density contributed by bright
stars (about 0.5\% of the critical density) and the BBN-determined mass
density of about 4.5\% of the critical density.  As this discussion
will illustrate, a baryon inventory is much easier to do at 1\,sec,
when the baryons exist as a smooth soup of hadronic matter,
than today, when they are dispersed in stars, stellar remnants,
hot gas, cold gas, and so on.

At redshifts of around 3 to 4, most of the baryons were still in
gas in the intergalactic medium (IGM).  This is what numerical simulations
of CDM say and what observations of the IGM at high redshift reveal.
At this time, structure was just beginning to form and can be observed
by studying the absorption of matter between us and distant quasars.
The baryon accounting based upon these observations does indeed
account for essentially all the baryons, though assumptions must
be made and the uncertainties are not as small as at BBN.\cite{weinberg}

In clusters of galaxies today the accounting is complete:  most of
the baryons are in the hot, intracluster gas that glows in x-rays.
The gas outweighs stars by about 10 to 1.  However, only about 5\%
of galaxies are in the great clusters of galaxies, so this leaves
the accounting very incomplete.  Globally, only about 1/3 of the
BBN baryon density can be accounted for, in the form of stars, cold
gas, and warm gas within galaxies.  The other 2/3 is {\em presumed} to be
in hot intergalactic gas and/or warm gas associated with galaxies.
One of the challenges for astrophysics is to complete the baryon
accounting today by detecting
this gas.  Efforts will involve both x-ray and UV instruments looking
for absorption or emission lines associated with the gas.

A dark horse possibility for the dark baryons is dark
stars (low-mass objects that never lit their nuclear fuels or the end
points of stellar evolution such as white dwarfs, neutron stars and
black holes that have exhausted their nuclear fuels).
Such objects in the halo of our own galaxy can be detected
by microlensing.  Microlensing of stars in the bulge of galaxy and in
the Large and Small Magallenic Clouds by dark, foreground objects
has been detected by the EROS, MACHO, DUO and OGLE groups.  This is
one of the exciting developments of the decade:  These rare (one in a
million or so stars is being lensed at any time) brightenings have
provided a new probe of the dark side of the Universe.
Already binary lenses, a
black-hole candidate, planets and important information about the structure
of the galaxy (strong evidence for a bar at the center) have been
revealed, and one very intriguing mystery remains.

While the handful of events toward the SMC can be explained as ``self
lensing,'' foreground objects in the SMC lensing SMC stars, the more than
twenty occurrences of microlensing of LMC stars are not so easily
understood.  Because the LMC is (thought to be) more compact, self lensing
is less important.  If one interprets the LMC lenses as a halo population
of dark objects, they would account for about 50\% of own halo.   The
mass inferred from the timescale of the brightenings (about 0.5\,$M_\odot$)
and the stringent limits to the number of main-sequence stars 
of this mass points to
white dwarfs.  (Recent HST observations give evidence for a handful
of nearby, fast-moving white dwarfs, consistent with a halo population of
white dwarfs.)

Beyond that, nothing else makes sense for
this interpretation.  Since white dwarf formation is very
inefficient there should
be 6 to 10 times as much gas left over as there are white dwarfs.
This of course would exceed the total mass budget of the halo by a wide
margin. The implied star formation rate
exceeds the measured star formation rate in the Universe
by more than an order of magnitude.  And where are their siblings who are still
on the main sequence?

Since microlensing only determines a line integral
of the density of lenses toward the LMC, which is
heavily weighed by the nearest 10 kpc or so, it gives little information about
where the lenses are.  Its limitations for probing the
halo are significant:  It cannot probe the halo at distances greater than
the distance to the LMC (50 kpc), and as a practical matter it can
only directly probe the innermost 15 kpc or so of the halo.  Recall,
the mass of the halo increases
with radius and the halo extends at least as far as 200 kpc.

Alternative explanations for the LMC lenses have been suggested.\cite{gg}
An unexpected component of the galaxy (e.g., a warped and
flaring disk, a very thick disk component, a heavier than expected spheroid,
or a piece of cannibalized satellite-galaxy between us and the LMC)
which is comprised of conventional objects (white dwarfs or lower-main sequence
stars); LMC self lensing (the LMC is being torn apart by the Milky
Way and may be more extended than thought); or a halo
comprised of 0.5\,$M_\odot$ primordial black holes formed around the time of
the quark/hadron transition (which also acts as the cold dark matter).
For all but the last, very speculative explanation,
the mass in lenses required is less than 10\% of the halo.

Because the cold dark matter framework is so successful and a baryonic halo
raises so many problems (in addition to those above, how to form large-scale
structure), I am putting my money on a CDM halo.  More data from microlensing
is crucial to resolving this puzzle.  The issue might also be settled by a
dazzling discovery:  direct detection of halo neutralinos or axions
or the discovery of supersymmetry at the Tevatron or LHC.

\section{Dark Energy}

The discovery of accelerated expansion in 1998 by the two supernova
teams (Supernova Cosmology Project and the High-z Supernova Team) was the
most well anticipated surprise of the century.  It may also be one
of the most important discoveries of the century.  Instantly, it
made even the most skeptical astronomers take inflation very seriously.
As for the hard-core, true-believers like myself, it suffices to
say that there was a lot of dancing in the streets.

\subsection{Anticipation}

In 1981 when Alan Guth put forth inflation most
astronomers responded by saying it was an interesting idea, but that its
prediction of a flat universe was at variance with cosmological fact.  At
that time astronomers argued that the astronomical evidence pointed
toward $\Omega_M \sim
0.05 - 0.10$ (even the existence of a gap between $\Omega_B$ and $\Omega_M$
was debatable).  Inflationists took some comfort in the fact that
the evidence was far from conclusive; it was largely
based upon the mass-to-light ratios of galaxies and clusters of galaxies,
and it did not sample sufficiently
large volumes to reliably determine the mean density of matter.  As
techniques improved, $\Omega_M$ rose.  Especially encouraging (to
inflationists) were the determinations of $\Omega_M$ based upon peculiar
velocity data (large-scale flows).  They not only probed larger
volumes and the mass more directly, but also by the early 1990s indicated
that $\Omega_M$ might well be as large as unity.

Even so, beginning in the
mid 1980s, the Omega problem ($\Omega_M \ll 1$) received much attention
from theorists who emphasized that the inflationary prediction was
a flat universe ($\Omega_0 = 1$), and not $\Omega_M =1$ (though certainly
the simplest  possibility).  A smooth, exotic component was suggested to
close the gap between $\Omega_M$ and 1 (smooth, so that it would not
show up in the inventory of clustered mass).  Possibilities discussed included
a cosmological constant (vacuum energy), relativistic particles produced
by the recent decay of a massive particle relic and a network of frustrated
topological defects.\cite{KT}

By 1995 it seemed more and more unlikely that
$\Omega_M =1$; especially damning was the determination of $\Omega_M$
based upon the cluster baryon fraction discussed earlier.
On the other hand, the
CDM scenario was very successful, especially if $\Omega_M h \sim 1/4$
(the shape of the power spectrum of density inhomogeneity today depends
upon this product because it determines the epoch when the Universe
becomes matter dominated).  Add to that the tension between the
age of the Universe and the Hubble constant, which is exacerbated for
large values of $\Omega_M$.  $\Lambda$CDM, the version of CDM with
a cosmological constant ($\Omega_M\sim 0.4$ and $\Omega_\Lambda
\sim 0.6$), was clearly the best fit CDM model (see Figure 2).
And it has a smoking gun signature:  accelerated expansion ($q_0 = {1\over 2}
-{3\over 2}\Omega_\Lambda$).  At the June 1996 Critical
Dialogues in Cosmology meeting at Princeton, in the CDM beauty
contest the only mark against
$\Lambda$CDM was the early result from the
Supernova Cosmology Project indicating that $\Omega_\Lambda
< 0.5\, (95\%)$.\cite{princeton}

\begin{figure}
\centerline{\psfig{figure=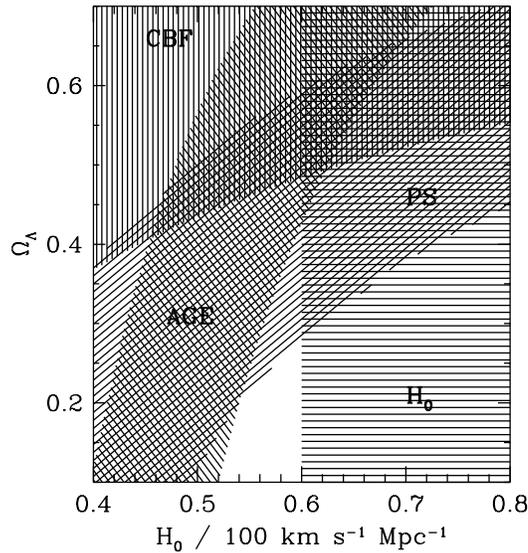,width=2.7in}}
\caption{Constraints used to determine the best-fit CDM model:
PS = large-scale structure + CBR anisotropy; AGE = age of the
Universe; CBF = cluster-baryon fraction; and $H_0$= Hubble
constant measurements.  The best-fit model, indicated by
the darkest region, has $H_0\simeq 60-65\,{\rm km\,s^{-1}
\,Mpc^{-1}}$ and $\Omega_\Lambda \simeq 0.55 - 0.65$.
}
\end{figure}

After the Princeton meeting the case grew stronger as CMB anisotropy
results began to define the
first acoustic peak at around $l=200$, as predicted in a flat Universe
(the position of the first peak scales $l=200/\sqrt{\Omega_0}$).  Today,
the data imply $\Omega_0=1\pm 0.1$\cite{kd} (see Figure 3).
With results from the Boomerang
Long-duration Balloon experiment expected in January, the DASI experiment
at the South Pole next summer, and the launch of the MAP satellite
in the Fall of 2000, we can expect a truly definitive determination
of $\Omega_0$ soon.

\begin{figure}
\centerline{\psfig{figure=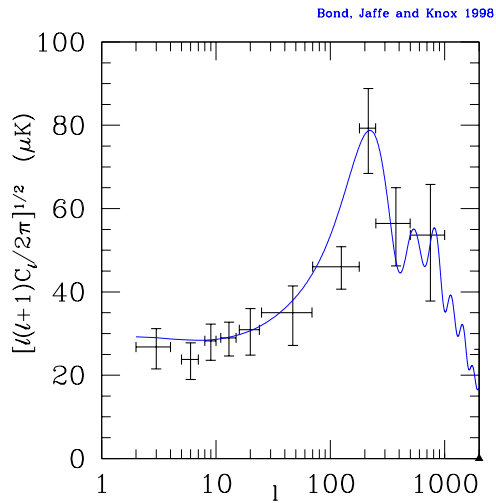,width=2.7in}}
\caption{Summary of CMB anisotropy measurements, binned to reduce error bars.
The theoretical curve is for the $\Lambda$CDM model with $H_0=65\,{\rm km\,
s^{-1}\,Mpc^{-1}}$ and $\Omega_M =0.4$ (Figure courtesy of L. Knox).
}
\end{figure}

The smoking-gun confirmation came in early 1998 with the
results from the two supernova groups indicating that the Universe
is speeding up, not slowing down.  Everything now fit together:
inflation and the flat universe,
the CMB determination that $\Omega_0\sim 1$ and the cluster
measurement of $\Omega_M\sim 0.4$, and the successes of CDM,
and $\Lambda$CDM in particular (see Figures 2-4).  In the minds
of theorists like me, the only surprise was that it took the
cosmological constant to make everything work.  Everything was
pointing in that direction, and were it not to the checkered
history of the cosmological constant, there would have been
no surprise at all.

\begin{figure}
\centerline{\psfig{figure=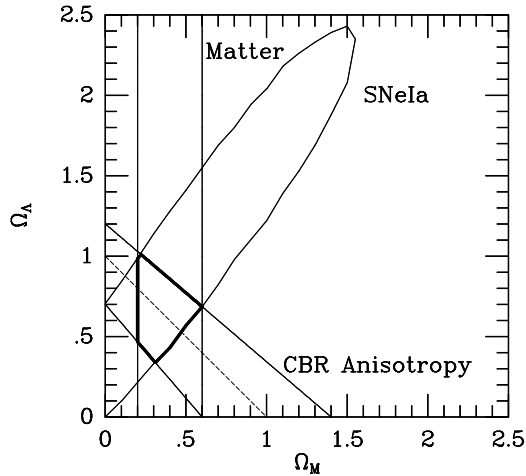,width=2.7in}}
\caption{Two-$\sigma$ constraints to $\Omega_M$ and $\Omega_\Lambda$
from CBR anisotropy, SNe Ia, and measurements of clustered matter.
Lines of constant $\Omega_0$ are diagonal, with a flat
Universe indicated by the broken line.
The concordance region is shown in bold:  $\Omega_M\sim 1/3$,
$\Omega_\Lambda \sim 2/3$, and $\Omega_0 \sim 1$.
}
\end{figure}

\subsection{The dark-energy problem}

At the moment, a crucial element in the case for accelerated expansion
and dark energy is the ``independent confirmation'' based
upon the otherwise discrepant numbers $\Omega_0\sim 1$ and $\Omega_M
\sim 0.4$.  Balancing the books requires a component
that is smooth and contributes about 60\% of the critical density.
In order that it not interfere with the growth of structure, its energy
density must evolve more slowly than matter so that there is a
long matter-dominated era during which the observed structure today
can grow from the density inhomogeneities measured by COBE and
other CMB anisotropy experiments.  Since
$\rho_X \propto R^{-3(1+w_X)}$, this places an upper limit to $w_X$:\cite{believe}
$w_X < -{1\over 2}$, and in turn, an upper limit to $q_0$:  $q_0<{1\over 2}
-{3\over 4}\Omega_X < 0$ for $\Omega_X > {2\over 3}$ and a
flat Universe.

Because of the checkered history of the cosmological constant --
cosmologists are quick to invoke it to solve problems
that later disappear and particle physicists have failed
to compute it to an accuracy of better than a factor of $10^{55}$ --
there is an understandable reluctance to accept it
without some skepticism.  To wit, other
possibilities have been suggested:  For example, a rolling scalar
field (essentially a mini-episode of inflation),\cite{roll} or a frustrated
network of very light topological defects (strings of walls).\cite{frustrate}

My preference is to characterize it as simply and most generally
as possible, by its equation of state:  $p_X=w_X\rho_X$, where
$w_X$ is $-1$ for vacuum energy, $-{N\over 3}$ for frustrated
topological defects of dimension $N$, and time-varying and between
$-1$ and $1$ for a rolling scalar field.  The goal then is to
determine $w_X$ and test for its time variation.\cite{xmatter}

In determining the nature of dark energy, I believe that telescopes and
not accelerators will play the leading role  -- even if there is
a particle associated with it, it is likely to be extremely difficult to
produce at an accelerator because of its gravitational or weaker
interactions with ordinary matter.  Specifically, I
believe that type Ia supernovae will prove to be the most powerful probe.
The reason is two fold:  first, the dark energy has only recently
come to be important; the ratio $\rho_M/\rho_X = (\Omega_M/\Omega_X)
(1+z)^{-3w_X}$ grows rapidly with redshift.  Secondly, dark energy
does not clump (or at least not significantly), so
its presence can only be felt through
its effects on the large-scale dynamics of the Universe.  Type Ia
supernovae have the potential of reconstructing the recent history of
the evolution of the scale factor of the Universe and from it, to
shed light on the nature of the dark energy.
Figures 5 and 6 show the simulated
reconstruction of two dark-energy models:
a quintessence model (scalar field rolling down a potential) by
means of supernovae and a variable equation of state.

Once one is convinced with high confidence
that there is dark energy out there (the next round of CMB anisotropy
results will be crucial) and that type Ia supernovae are standardizable
candles (more study of nearby supernovae), the next step is a
dedicated assault, probably a satellite based telescope
(which I like to call DaRk-Energy eXplorer or D-REX) to collect
1000s of supernovae between redshift 0 and 1.  By carefully culling the
sample and doing good follow up and accurate photometry one will able to
address determine $\Omega_X$, $w_X$ and address the time variation
of $w_X$.\cite{dragan}

\begin{figure}
\centerline{\psfig{figure=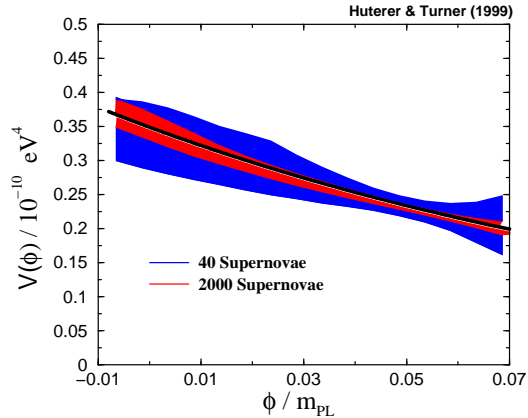,width=2.7in}}
\caption{The 95\% confidence bands for the simulated reconstruction
of the scalar-field potential for a quintessence model; the solid
curve is the original potential.
}
\end{figure}

\begin{figure}
\centerline{\psfig{figure=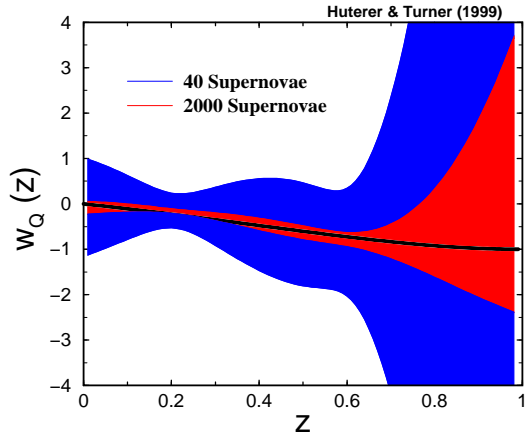,width=2.7in}}
\caption{The 95\% confidence bands for the simulated reconstruction
of the equation of state, $w(z)$; the solid
curve is the input equation of state.  Note, because dark energy
becomes less important (relative to matter) at high redshift,
it becomes more difficult to probe its properties.
}
\end{figure}

\section{Looking Forward}

The two dominant ideas in cosmology over the past 15 years have been
inflation and cold dark matter.  They have provided the field with a
grand guiding paradigm which has spurred the observers and
experimenters to put in place a remarkable program that keep the
field of cosmology lively for at least two decades.

Over the past few years this paradigm has begun to be tested in a
significant way, with many more important tests to come.  The first tests
have been encouraging.  The first acoustic peak in the CMB power spectrum
indicates a flat Universe and is consistent with the scale-invariant
inflationary power spectrum which predicts a series of acoustic peaks.
The discovery of accelerated expansion provided the evidence for the
component that balanced the books:
our flat universe = 40\% dark matter + 60\% dark energy.
This is only the beginning of this great adventure.

Central to cosmology and the connection between cosmology and
fundamental physics are the two dark problems:  dark matter and
dark energy.  The dark matter problem is more than sixty years old
and quick mature.  We have divided the dark matter problem into
two distinct problems, dark baryons and nonbaryonic dark matter, and
narrowed the possibilities for each.  The baryons are most likely
in the form of diffuse, hot gas.  The nonbaryonic dark matter is most
likely slowly moving, essentially noninteracting particles
(cold dark matter), with relic
elementary particles from the earliest moments being the leading
candidate.  Foremost among them are the axion and neutralino.

We could still be in for some surprises:  the CDM particles could be
something more exotic (primordial black holes or superheavy particles
produced in the reheating process at the end of inflation).  Likewise, the
simple and thus far very successful assumption that the only interactions
of the CDM particles that are relevant today are gravitational, could
be wrong.  There are some hints otherwise:  The halo profiles
predicted for noninteracting CDM are too cuspy at the center.\cite{Moore}
The resolution could be astrophysical or it could involve fundamental physics.
Perhaps, it is indicating that the CDM particles have significant
interactions today (scattering or annihilations) that round off the central
cusps.  It is intriguing to note that neither the axion nor the neutralino
has such interactions.

By comparison, the dark-energy problem is
in its infancy.  The evidence for it, while solid, is not air tight.
Unlike the dark-matter problem where sixty years of detective work have
brought us to a couple of very specific suspects, the
possibilities for the dark energy are wide open.
But two things are clear:  as with the dark-matter problem, the solution
certainly involves fundamental physics, and telescopes will play a major role
in clarifying the nature of the dark energy.

\section*{Acknowledgments}
This work was supported by the DoE (at Chicago and Fermilab)
and by the NASA (through grant NAG 5-7092 at Fermilab).


\section*{References}
\begin{thebibliography}{99}

\bibitem{H0}  See e.g., J.R. Mould et al, astro-ph/9909260.

\bibitem{SNe}  A.G. Riess, et al., {\em Astron. J.} {\bf 116}, 1009 (1998);
S. Perlmutter et al, {\em Astrophys. J.} {\bf 517}, 565
(1999) (astro-ph/9812133).

\bibitem{DM_SUM} See e.g., M.S. Turner, astro-ph/9811454.

\bibitem{krauss} L. Krauss and M.S. Turner, {\em Gen. Rel. Grav.}, in
press (astro-ph/9904020).

\bibitem{IAU}  {\em Dark Matter in the Universe} (IAU Symposium \#117),
eds. J. Knapp and J. Kormendy (Reidel, Dordrecht, 1987).

\bibitem{BBN}  S. Burles et al., {\em Phys. Rev. Lett.} {\bf 82},
4176 (1999).

\bibitem{CBF}  J. Mohr et al, {\em Astrophys. J.}, in press (1999)
(astro-ph/9901281).

\bibitem{candidates} B. Sadoulet, {\em Rev. Mod. Phys.} {\bf 71},
S197 (1999).

\bibitem{SuperK} Y. Fukuda et al, {\em Phys. Rev. Lett.} {\bf 81},
1562 (1998).

\bibitem{weinberg} See e.g., D. Weinberg et al, {\em Astrophys. J.}
{\bf 490}, 546 (1997).

\bibitem{gg}  E. Gates and G. Gyuk, astro-ph/9911149.

\bibitem{KT}  E.W. Kolb and M.S. Turner, {\em The Early Universe}
(Addison-Wesley, Redwood City, CA, 1990).

\bibitem{kd}  S. Dodelson and L. Knox, astro-ph/9909454.

\bibitem{princeton}  {\em Critical Dialogues in Cosmology}, ed.
N. Turok (World Scientific, Singapore, 1997);
L. Krauss and M.S. Turner, {\em Gen. Rel. Grav.}
{\bf 27}, 1137 (1995).

\bibitem{believe} M.S. Turner, astro-ph/9904049.

\bibitem{roll} M. Bronstein, {\em Phys. Zeit. Sowjet Union}
{\bf 3}, 73 (1933);
M. Ozer and M.O. Taha, {\em Nucl. Phys. B} {\bf 287} 776 (1987);
K. Freese et al., {\em ibid} {\bf 287} 797 (1987);
L.F. Bloomfield-Torres and I. Waga, {\em Mon. Not. R. astron. Soc.}
{\bf 279}, 712 (1996); J. Frieman et al, {\em Phys. Rev. Lett.}
{\bf 75}, 2077 (1995); K. Coble et al, {\em Phys. Rev. D}{\bf 55},
1851 (1996); R. Caldwell et al, {\em Phys. Rev. Lett.} {\bf 80},
1582 (1998); B. Ratra and P.J.E. Peebles, {\em Phys. Rev. D}
{\bf 37}, 3406 (1988).

\bibitem{frustrate} A. Vilenkin, {\em Phys. Rev. Lett.} {\bf 53}, 1016 (1984);
D. Spergel and U.-L. Pen,  {\em Astrophys. J.} {\bf 491}, L67 (1997).

\bibitem{xmatter} M.S. Turner and M. White, {\em Phys. Rev. D}
{\bf 56}, R4439 (1997); S. Perlmutter, M.S. Turner and M. White,
{\em Phys. Rev. Lett.} {\bf 83}, 670 (1999).

\bibitem{dragan}  D. Huterer and M.S. Turner, {\em Phys. Rev. D}
{\bf 60}, 081301 (1999); and in preparation.

\bibitem{Moore} B. Moore, {\em Nature} {\bf 370}, 629 (1994).




\end{thebibliography}
\end{document}